\documentclass[aip,apl,reprint,showpacs,superscriptaddress,groupedaddress]{revtex4-1}  
\usepackage{graphicx}  
\usepackage{dcolumn}   
\usepackage{bm}        
\usepackage{amssymb}   
\usepackage{color}

\begin{document}

\title{Spin field-effect transistor action via tunable polarization of the spin injection in a Co/MgO/graphene contact}
\author{Sebastian Ringer} \affiliation{Institute of Experimental and Applied Physics, University of Regensburg, D-93040 Regensburg, Germany}
\author{Matthias Rosenauer} \affiliation{Institute of Experimental and Applied Physics, University of Regensburg, D-93040 Regensburg, Germany}
\author{Tobias V\"olkl} \affiliation{Institute of Experimental and Applied Physics, University of Regensburg, D-93040 Regensburg, Germany}
\author{Maximilian Kadur} \affiliation{Institute of Experimental and Applied Physics, University of Regensburg, D-93040 Regensburg, Germany}
\author{Franz Hopperdietzel} \affiliation{Institute of Experimental and Applied Physics, University of Regensburg, D-93040 Regensburg, Germany}
\author{Dieter Weiss} \affiliation{Institute of Experimental and Applied Physics, University of Regensburg, D-93040 Regensburg, Germany}
\author{Jonathan Eroms} \affiliation{Institute of Experimental and Applied Physics, University of Regensburg, D-93040 Regensburg, Germany}
\email{jonathan.eroms@ur.de}
    
\date{\today}

\begin{abstract}
We fabricated a non-local spin valve device with Co-MgO injector/detector tunnel contacts on a graphene spin channel. In this device, the spin polarization of the injector contact can be tuned by both the injector current bias and the gate voltage. The spin polarization can be turned off and even inverted. This behavior enables a spin transistor where the signal is switched off by turning off the spin injection using the field-effect.	We propose a model based on a gate-dependent shift of the minimum in the graphene density of states with respect to the tunneling density of states of cobalt, which can explain the observed bias and gate dependence.

\end{abstract}

\maketitle

Spintronics expands electronics from using only the charge of the electron to also utilizing its spin property \cite{Zutic2004}. So far, spintronic devices have only been used for data storage, but concepts exist for also building spin based logic circuitry \cite{DattaDas1990, Behin-Aein2010, Dery2012}. The paradigmatic device, the spin field effect transistor was proposed by Datta and Das in 1990 \cite{DattaDas1990}. Here, spins are injected from a ferromagnetic electrode into a non-magnetic channel, and a spin-dependent signal is detected at a second, ferromagnetic electrode. Spins are rotated along the way by a gate-tunable spin-orbit interaction. While this device allows for an all-electric control of spin signals, in contrast to magnetic switching of the electrodes, the channel needs to have strong and tunable spin-orbit interaction and, at the same time, a long spin lifetime. Because of these conflicting requirements, an attempt to fully realize the Datta-Das transistor was only presented more than two decades after the original proposal \cite{Koo2009}. On the other hand, when the  polarization of spin injection can be manipulated electrically, a transistor device can be realized in a device with long spin lifetime in the channel, such as graphene \cite{Tombros2007}. Electric tunability of spin injection has been demonstrated in magnetic tunnel junctions \cite{Sharma1999, DeTeresa1999, Tiusan2007, Godel2014, Asshoff2017}, {or Si based devices \cite{Dankert2013}.}
For graphene spintronics, bias-dependent spin polarization{, including signal inversion,} was reported for hexagonal boron nitride (hBN) \cite{Kamalakar2016,Gurram2017} or MgO \cite{Han2012,Han2009} as a tunnel barrier. {In a MoS$_2$/graphene heterostructure, a gate-dependent suppression of the spin signal was reported \cite{Dankert2017}.} 
However, no gate-controlled signal inversion was reported for tunneling spin injection in graphene devices.

{In this work, we report on a gate and bias-tunable spin polarization in a Co/MgO/graphene device. Importantly, the sign of spin polarization can be reversed, and spin injection can even be turned off by gate control, thus enabling a true three-terminal spintronic device.}
{We find that} at an elevated negative injector bias $U_{inj}$, the spin polarization vanishes and afterwards changes sign. 
{At this bias setting, which we call spin neutrality point, both sign and magnitude of spin polarization can be controlled by a voltage $V_g$ applied to the back gate of the sample.}

\begin{figure}
\includegraphics[width=0.35\textwidth]{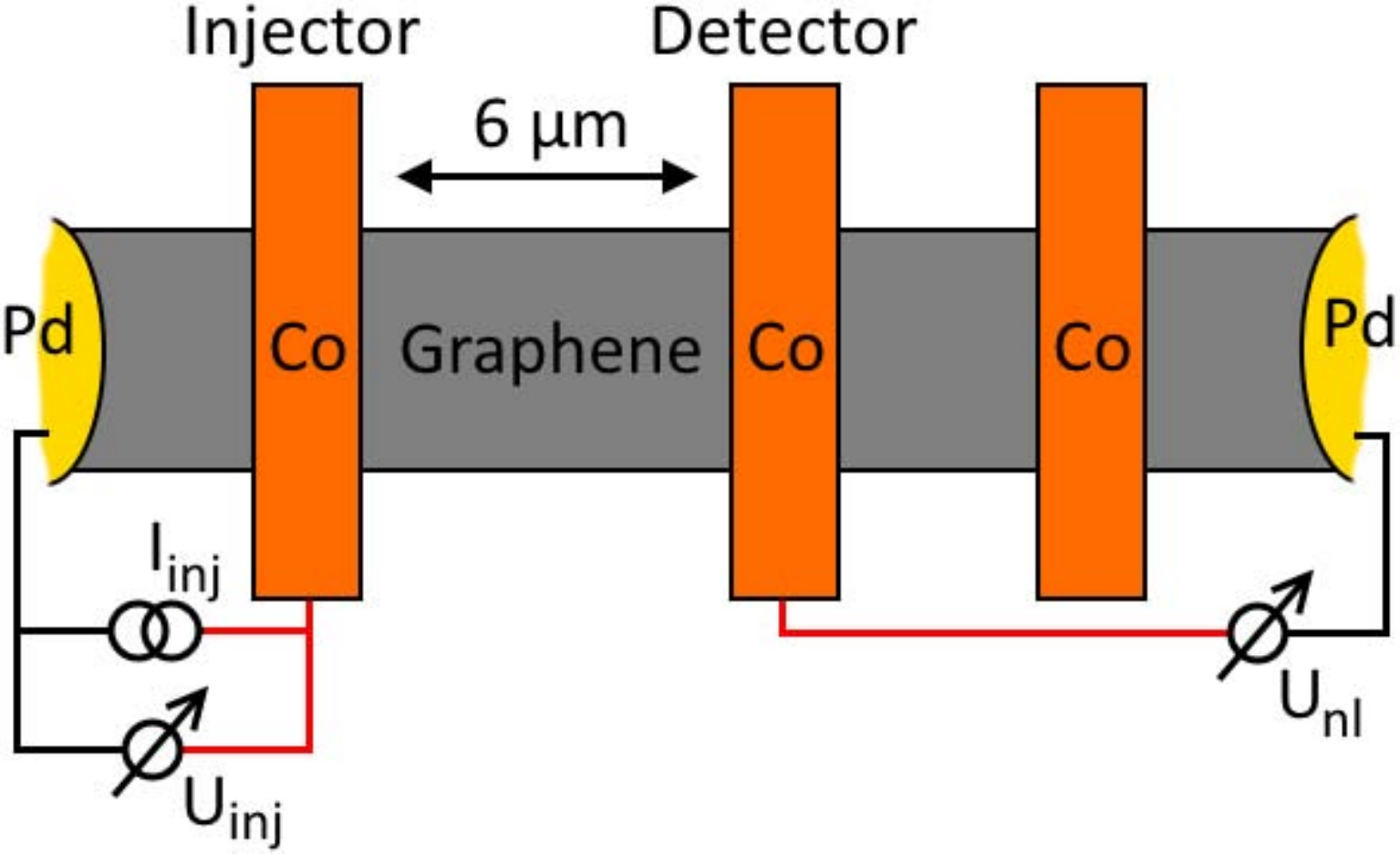}
\caption{\label{fig:sample}Sample schematic showing a graphene flake with contacts in the non-local spin valve measurement setup. {The polarity of current source and voltage detectors is indicated by red (positive) and black (negative) leads.}}
\end{figure}
\begin{figure}
\includegraphics[width=0.35\textwidth]{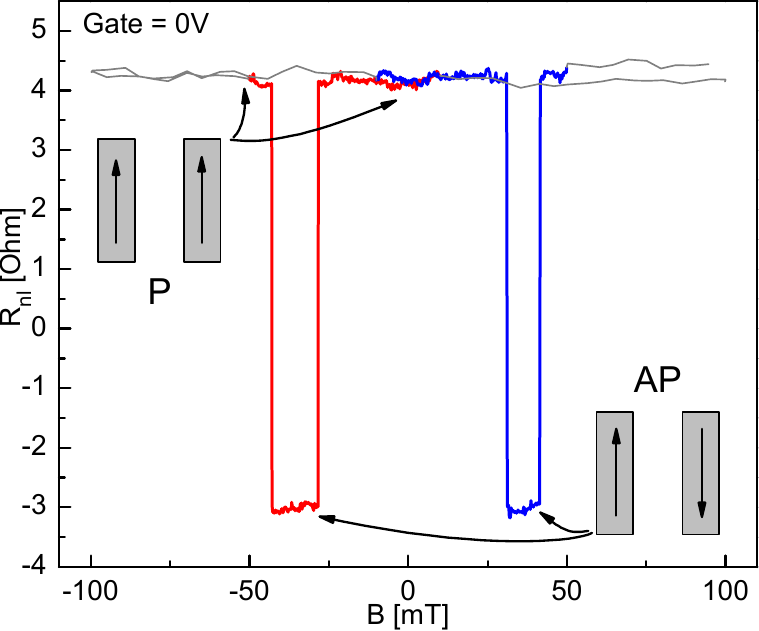}
\caption{\label{fig:spinvalve} Spin valve signal at $V_g = 0$\,V with illustrations to show the parallel (P) and antiparallel (AP) magnetic orientation of the electrodes. Distance of the injector and detector contacts was 6\,$\mu$m with an injector current of $I_{inj}=+5\,\mu$A. The grey trace shows the preparation of the electrodes that was done at a higher sweep rate, which induces a slight inductive offset because of the DC measurement setup.}
\end{figure}
\begin{figure}
\includegraphics[width=0.45\textwidth]{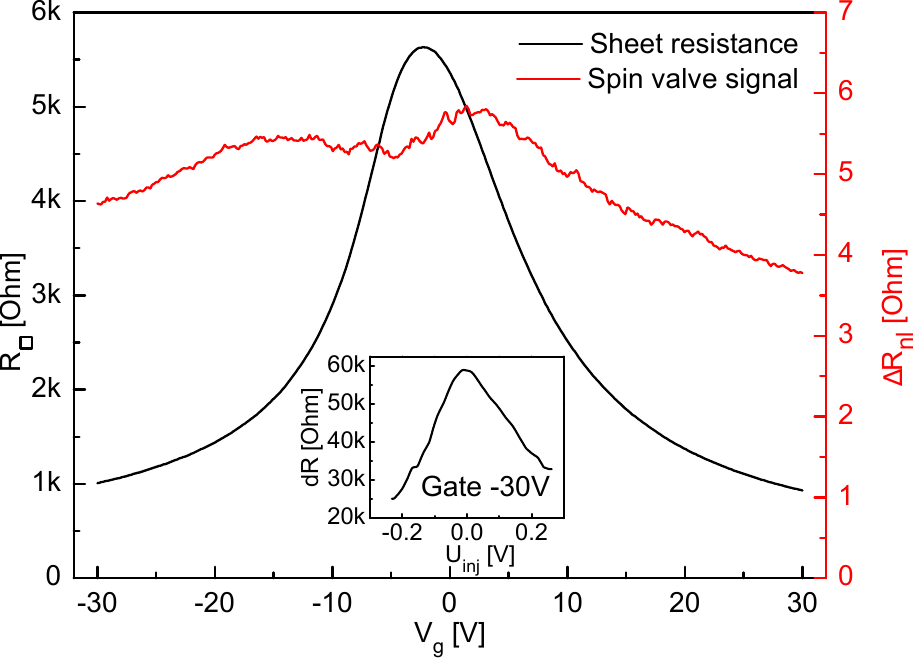}
\caption{\label{fig:gate} Gate dependence of the non-local spin valve signal $\Delta R_{nl}$ with an injector current of $I_{inj}=+4\,\mu$A (red) and the sheet resistance (black). The Dirac point is at $V_g=-2$\,V. Inset shows the differential resistance of the injector tunnel contact as a function of voltage bias.}
\end{figure}
\begin{figure}
\includegraphics[width=0.44\textwidth]{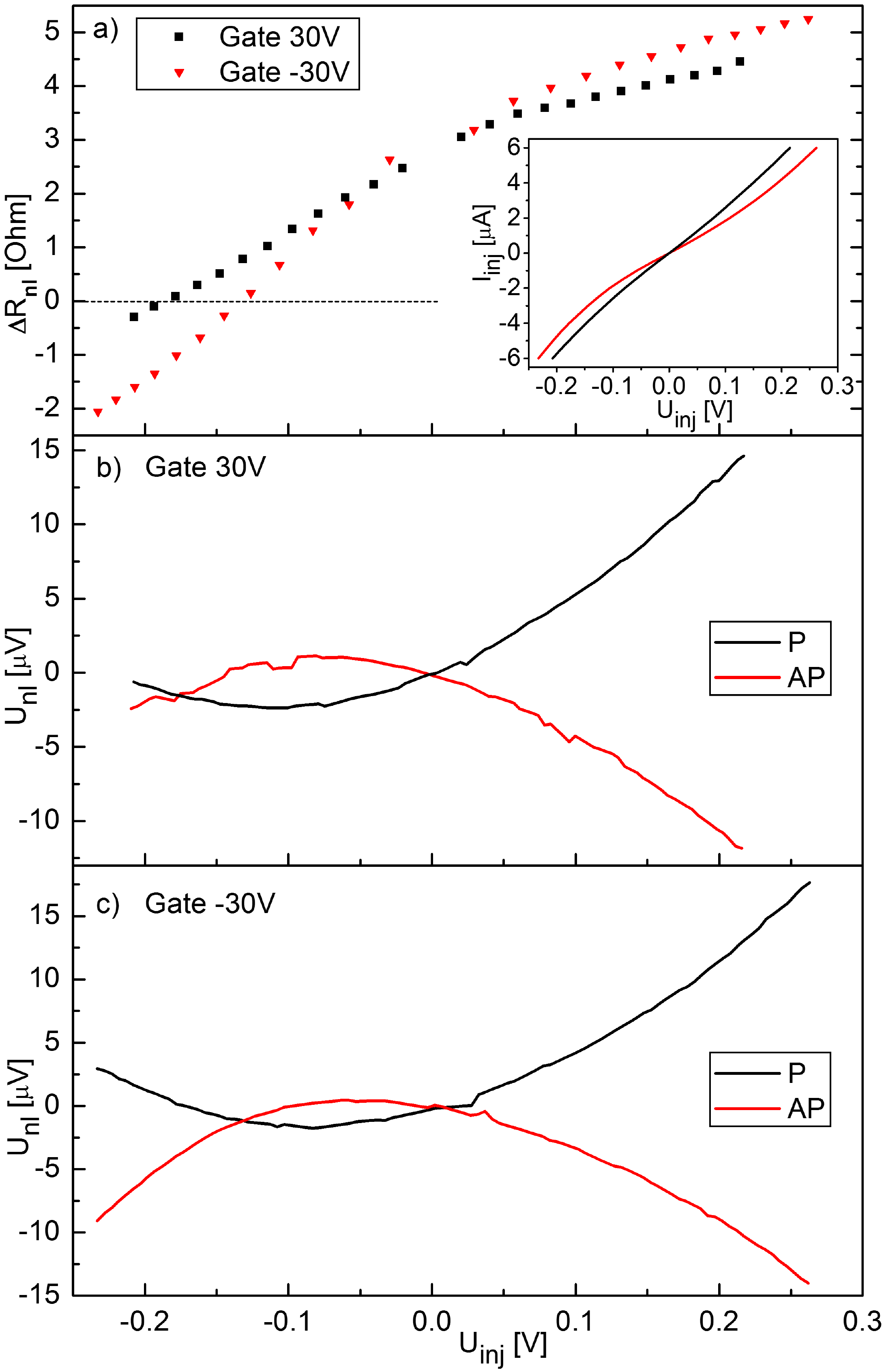}
\caption{\label{fig:Invert1} a) Plot of the spinvalve signal $\Delta R_{nl}$ in dependence of the injector bias for $V_g=30$\,V (black squares) and $V_g=-30$\,V (red triangles). Negative values indicate an inverted spin valve. The injector current was varied between 0.5 and 6\,$\mu$A with steps of 0.5\,$\mu$A for each current polarity and gate voltage. b) \& c) Plot of the bias measured at the detector electrode in dependence of the injector bias for parallel (P, black) and antiparallel (AP, red) configuration, for gate voltages of b) $V_g=30$\,V and c) $V_g=-30$\,V.}
\end{figure}

Fig. \ref{fig:sample} shows a schematic picture of the graphene flake and the non-local measurement setup. We use exfoliated single layer graphene on $\text{p}^{++}\text{Si/SiO}_2$, {where the highly doped silicon serves as a back gate, using the 285 nm thick SiO$_2$ as a dielectric.} The inner contacts are Co with $\sim$1.4\,nm of MgO as a tunnel barrier, the end contacts are Pd. More details on fabrication and spin transport properties of this device can be found in Ref. \onlinecite{Ringer2018}. At the injector electrode, a DC current bias $I_{inj}$ is applied and the bias voltage $U_{inj}$ is simultaneously measured. We use the well-known non-local detection scheme \cite{JohnsonSilsbee1985} to record a non-local voltage signal $U_{nl}$ and the corresponding non-local resistance $R_{nl}=
U_{nl}/I_{inj}$ at the detector electrode. We define the outer Pd electrode as ground for the injector circuit. A positive $U_{inj}$ or $I_{inj}$ therefore means an electron current flowing from graphene to the Co electrode. Similarly, the detector circuit is connected with the positive terminal of the nanovoltmeter to the Co detector electrode. Since we are using a DC setup, the bias dependence can be studied in this experiment, at the expense of sensitivity to unavoidable magnetic induction signals. The spin valve signal $\Delta R_{nl}$ is defined as $\Delta R_{nl}=R_{nl,P}-R_{nl,AP}$, for $R_{nl}$ in the parallel (P) or antiparallel (AP) configuration. 
All measurements were performed at $T=200$\,K in a cryostat equipped with a 3D vector magnet.
At a distance between the injector and the detector contacts of 6\,$\mu$m, we achieve a spin valve signal of $R_{nl}\sim7\,\Omega$ at an injector current of $I_{inj} = +5\,\mu$A, as can be seen in Fig.  \ref{fig:spinvalve}. 
The resistance area products of the injector and detector electrodes are 45.9\,k$\Omega\mu\text{m}^2$ and 27.0\,k$\Omega\mu\text{m}^2$, respectively. In Fig. \ref{fig:gate} we see in black a back gate sweep of the graphene sheet resistance, with the Dirac point at $V_g=-2$\,V. From these data, the carrier mobility was calculated to be between $\mu=3500 \dots 5000\, \text{cm}^2/\text{Vs}$, depending on back gate voltage. The red trace in Fig. \ref{fig:gate} displays the gate dependence of the spin valve signal, at an injector current of $I_{inj}=+4\,\mu$A. At this injector current the gate dependence follows qualitatively the shape for high quality tunnel contacts as described by W. Han \textit{et al.}\cite{Han2010}. The differential resistance $dR = \frac{dU_{inj}}{dI_{inj}}$ of the injector contact is shown in the inset of Fig. \ref{fig:gate} and displays clearly non-ohmic behavior, another indicator for high quality tunnel barriers.\\
\begin{figure}
\includegraphics[width=0.45\textwidth]{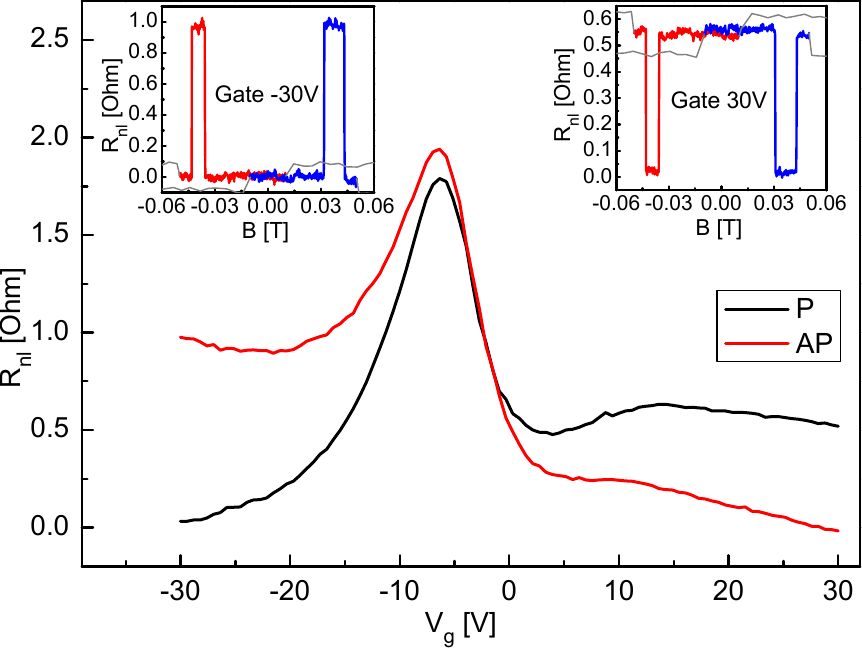}
\caption{\label{fig:Invert2} Gate dependence of the non-local resistance of parallel (P, black) and antiparallel (AP, red) configuration, at a fixed injector current of $I_{inj}=-4\,\mu$A. Insets show the spinvalve signal at $V_g=30$\,V and $V_g=-30$\,V.}
\end{figure}
\begin{figure}
\includegraphics[width=0.4\textwidth]{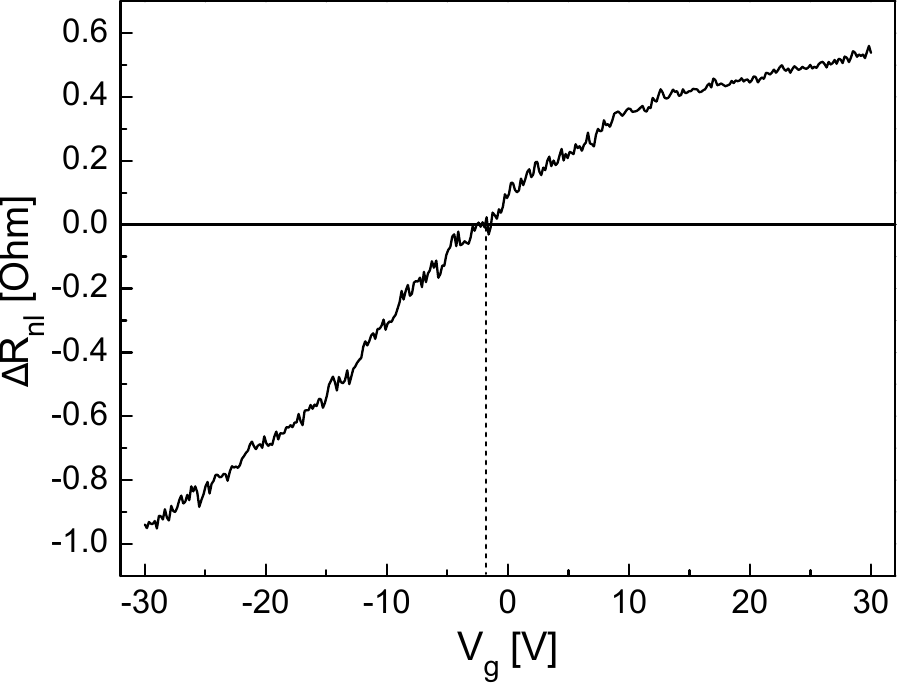}
\caption{\label{fig:transistor} Gate dependence of the non-local spin valve signal $\Delta R_{nl}$ at a fixed injector current of $I_{inj}=-4\,\mu$A in black.}
\end{figure}
Now we apply a negative bias current of $I_{inj}=-4\,\mu$A. Surprisingly, the observed polarity of the spin valve signal now depends on the gate voltage $V_g$, as shown in Fig. \ref{fig:Invert2} (insets). When we fix the electrode magnetization to either P or AP and record the gate reponse of $R_{nl}$, we observe the black (P) and red (AP) curves in Fig. \ref{fig:Invert2}. The traces cross, which indicates that the back gate can change the polarity of the detected spins. \\
The transition between these two states is best observed in Fig. \ref{fig:transistor} which displays the gate dependence of the non-local spin valve signal $\Delta R_{nl}$, calculated from the data in Fig. \ref{fig:Invert2}. A positive $\Delta R_{nl}$ then represents the expected spin valve signal while a negative $\Delta R_{nl}$ signifies an inverted spin valve signal. As can be seen, the transition between regular and inverted spin signal occurs at $V_g=-2$\,V. The transition is continuous and approximately linear. That the Dirac point is at the same back gate voltage as the spin neutrality point is a coincidence. 
Applying a different injector current will move the spin neutrality point, as shown below. 
As Fig. \ref{fig:transistor} shows, at negative bias the sample acts as spin field effect transistor, where the back gate can be used to turn $\Delta R_{nl}$ on, off or invert it.\\
Finally, we study the bias dependence of the spin valve signal. 
Fig. \ref{fig:Invert1}a) displays the dependence of $\Delta R_{nl}$ on the injector bias $U_{inj}$ at a gate voltage of $V_g=30$\,V (black squares) and $V_g=-30$\,V (red triangles). The inset shows the $U_{inj}$-$I_{inj}$ dependence of the injector contact for the corresponding gate voltages. Fig. \ref{fig:Invert1}b) and c) display the direct response of the detector voltage $U_{nl}$ to the injector bias $U_{inj}$, for $V_g = 30$\,V in Fig. \ref{fig:Invert1}b) and $V_g = -30$\,V in Fig. \ref{fig:Invert1}c). 
As Fig. \ref{fig:Invert1}a) shows, for positive $U_{inj}$, $\Delta R_{nl}$ is always positive, while at negative $U_{inj}$, $\Delta R_{nl}$ changes sign at a certain value of $U_{inj}$. This inversion point can be tuned by the gate voltage, or, equivalenty, an injector bias of $U_{inj}\approx -150$\,mV sets the operating point for gate-controlled spin transistor action.
As the measurements are performed in a lateral spin valve geometry, we need to distinguish between contributions from the spin channel and contributions from the tunnel contacts. As the spin valve inversion is only enabled by a negative injector current, this is a clear indicator of a contact effect. However, when applying a back gate voltage, contributions from the spin channel to the amplitude of $\Delta R_{nl}$ cannot be excluded. We note that the gate dependence of the spin injection polarization can pose a problem when a measurement of $\Delta R_{nl} (V_g)$ is used to extract spin transport parameters of the graphene channel \cite{Jozsa2009, Zomer2012}.\\
What is the mechanism behind the observed reversal of the spin polarization? The reversal of the spin polarization is originating from a property of the injector contact and appears at an injector bias of $U_{inj} \approx -150$\,mV. The ability to manipulate the effect by the back gate points to a connection to the Fermi level in graphene. For low-resistance contacts, the Fermi level in graphene under the electrodes is pinned{, {\em i.e.}, the Fermi level is influenced by the presence of Co electrodes and does not respond to a gate voltage} \cite{Volmer2013}. This pinning is lifted when the contact resistance is high enough, which we assume to be the case in our sample. {For a more in-depth discussion of Fermi level pinning, see Ref. \onlinecite{Volmer2013}.}
 A pinned Fermi level under the electrodes might be the reason for an absence of a {gate-dependent signal reversal} in other publications\cite{Han2009,Han2012}.\\
One mechanism that can lead to a strong energy dependence of the spin polarization are band structure effects that can be found in fully crystalline magnetic tunnel junctions \cite{Heiliger2005}. While possible in our case, we consider this mechanism unlikely as in our samples the MgO/Co layers are polycrystalline. A further possible origin are defects in the barrier, where resonant tunneling at a specific energy can result in  spin valve inversion \cite{Tsymbal2003}. This does not match with the featureless bias dependence in our sample as shown in Fig. \ref{fig:Invert1}.\\
\begin{figure*}
\includegraphics[width=0.75\textwidth]{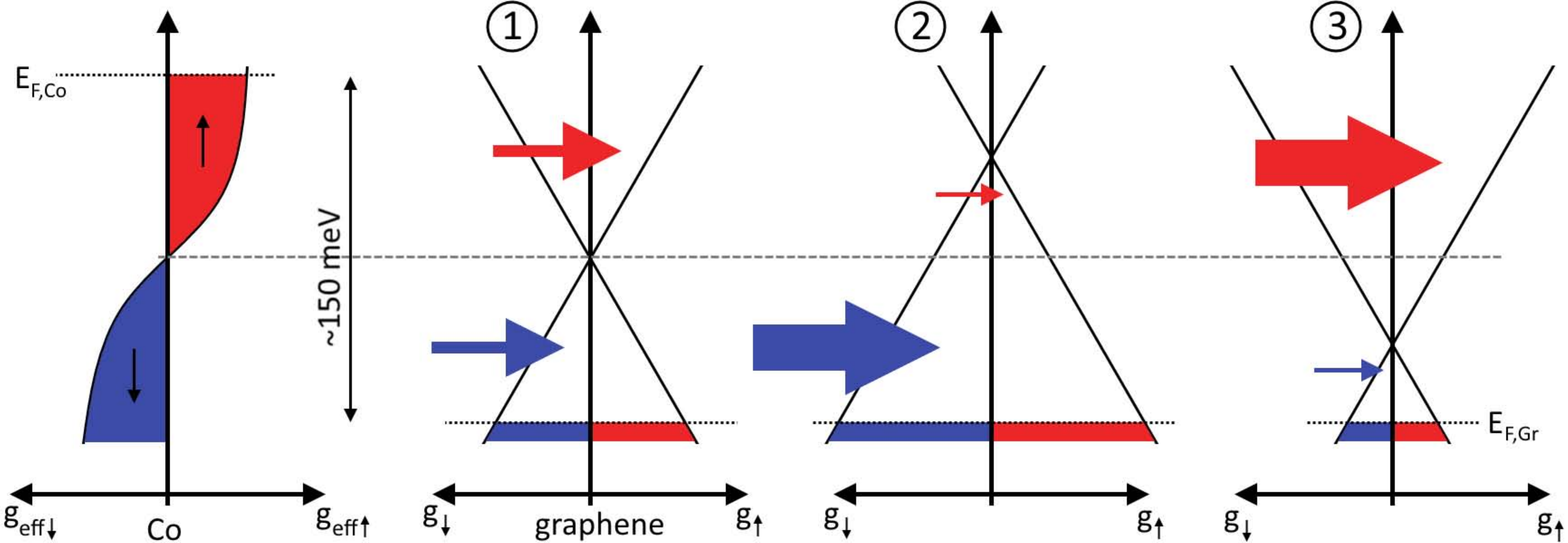}
\caption{\label{fig:spin-inversion-DOS} Proposed mechanism of the spin transistor action at an applied injector bias of $\sim -150$\,mV, showing the tunneling of electrons from Co to graphene. (1) shows the spin neutrality point, where injector bias, gate and doping align the tunneling DOS of Co and graphene to have equal tunneling of spin up and down electrons. The gate can then be used to move the Dirac point up or down on the energy scale, resulting in a spin down polarization in graphene in (2) and a spin up polarization in (3).}
\end{figure*}
When in contact with a ferromagnet, the band structure of graphene can be spin split through a magnetic proximity effect \cite{Lazic2016}. This can then lead to an inverted spin valve when tunneling from p to n doped graphene, as observed by Asshoff \textit{et al.} \cite{Asshoff2017} and Xu \textit{et al.} \cite{Xu2018}. However, due to the thick MgO barrier in our case -- as evidenced by the non-ohmic differential resistance shown in Fig.~\ref{fig:gate} -- there is no region of direct contact between Co and graphene, making this explanation unlikely. Finally, we discuss the spin polarization of the Co injectors. While the polarization $P_N$ derived from the spin-resolved density of states (DOS) of cobalt \cite{Blaha1988} stays constant in an energy window of about 700 meV below the Fermi level, for spin injection one has to consider the spin polarization weighted by $v^\alpha$, where $v$ is the electron velocity, and $\alpha=1$ or 2 for a ballistic or diffusive situation, respectively \cite{Mazin1999}. Mazin calculated this quantity for Fe and Ni, showing a sign change of $P_{Nv^2}$ around the Fermi level for Nickel, fundamentally different from $P_N$ of Ni, while for Fe, both $P_{Nv^2}$ and $P_N$ show similar behavior around $E_F$ \cite{Mazin1999}. The quantities $P_N$, $P_{Nv}$ and $P_{Nv^2}$ were calculated for Co/graphene by Sipahi \textit{et al.} \cite{Sipahi2014}. They observe that, while $P_N$ in the bulk Co layers retains its sign over a wide energy range, the velocity-weighted polarization can show a strong energy dependence, making this a possible explanation for the observed effect. Furthermore, the spin polarization can be quite different at the interface, depending on its detailed conditions \cite{Moser2006}. Lou {\it et al.} reported a strong bias dependence in a Fe/GaAs spin valve device, offering the band structure at the Fe/GaAs as a possible explanation \cite{Lou2007}. For our situation, the relevant interface is the Co/MgO interface. Unfortunately, due to the unknown crystal structure in our non-epitaxial samples, a comparison to first-principles calculations \cite{Kim2008} is not meaningful.\\

Given that reversal of spin polarization was reported in Co-graphene devices with both MgO\cite{Han2012,Zhu2018} and hBN tunnel barriers\cite{Gurram2017,Kamalakar2016,Zhu2018} for negative injector bias in the range of order 100 mV, we assume that indeed the effective spin polarization $g_{eff}$ of the tunnel current from Co into graphene has a sign change slightly below the Fermi level, as shown in Fig. \ref{fig:spin-inversion-DOS}. 
{While the uncovered graphene regions show almost no doping (cf. resistance curve in Fig~\ref{fig:gate}), both the difference in work function of Co and graphene and the electric field induced by the bias voltage can lead to a different position of the Fermi level underneath the Co electrodes.}
Assuming p-doping of the graphene underneath the contacts, an applied bias voltage of $\sim -150$\,mV then leads to equal amounts of electrons with negative and positive spin polarization entering graphene, thus cancelling the spin signal (situation (1) in Fig. \ref{fig:spin-inversion-DOS}). When a gate voltage is applied, the position of the Dirac point in graphene is changed, while the bias voltage fixes the separation of the Fermi levels in Co and graphene, respectively. Since the DOS in graphene strongly changes around the Dirac point, shifting the Dirac cone up or down will select either preferentially down spins (situation (2)) or up spins (situation (3) in Fig. \ref{fig:spin-inversion-DOS}), leading to the observed inversion of the spin signal. The signal is inverted for a larger proportion of down spins entering graphene, which is favored by a Fermi level lying deeper in the valence band of graphene ({\em i.e.}, negative gate voltage), as found in the experiment ({\em cf.} Fig. \ref{fig:transistor}).
The precise position of the contact-induced doping sets the bias value of the spin neutrality point, which can explain its variation in samples prepared in different groups. 
{For zero bias, the Fermi levels in graphene and Co are aligned, and the spin polarization of the tunneling current is only probed close to $E_F$. As the DOS of Co is fairly constant around $E_F$, changing the Fermi level in graphene by a gate voltage has little effect on the spin signal.}

In conclusion, we report on a tunable spin polarization of injected spins through a \mbox{Co/MgO/graphene} contact. For a certain range of negative injector bias the spin polarization can be controlled by the back gate, turning the device into a spin field effect transistor. We consider an energy dependence of the spin polarization at the Co/MgO interface as the most likely cause for the observed effect and propose a model that offers a conclusive explanation. In addition to the possible application as a spin transistor, we note that the gate dependence of the spin polarization has to be taken into account when studying the correlation of the spin lifetime to other gate dependent parameters of graphene.

During the preparation of this manuscript, Zhu {\it et al.} \cite{Zhu2018} published a systematic study of the bias-dependent spin polarization reversal in Co/MgO/graphene and Co/hBN/graphene tunnel junctions. They did investigate the gate dependence, but did not observe an effect of comparable magnitude as in our sample.

\begin{acknowledgments}
Financial support by the Deutsche Forschungsgemeinschaft (DFG) within the programs SFB 689 and GRK 1570 is gratefully acknowledged, as well as financial support by the Elitenetzwerk Bayern. The authors would like to thank F. Volmer for crucial advice on improving the quality of our MgO tunnel barriers, and K. Zollner and J. Fabian for discussions. 
\end{acknowledgments}


%

\end{document}